\documentclass{aa}

\usepackage{graphicx,amssymb,natbib}
\usepackage{txfonts}
\usepackage{mathrsfs}
\usepackage{color}

\begin{document}

\title{First simultaneous SST/CRISP and IRIS observations of a small-scale quiet Sun vortex}

\author{S.-H. Park\inst{1}\fnmsep\thanks{Present address: Trinity College Dublin, College Green, Dublin 2, Ireland}, G. Tsiropoula\inst{1}, I. Kontogiannis\inst{1}, K. Tziotziou\inst{1}, E. Scullion\inst{2}, \and J.G. Doyle\inst{3}}

\institute{Institute for Astronomy, Astrophysics, Space Applications and Remote Sensing (IAASARS), National Observatory of Athens, Penteli 15236, Greece; shpark@noa.gr \and Trinity College Dublin, College Green, Dublin 2, Ireland \and Armagh Observatory, College Hill, Armagh BT61 9DG, N. Ireland}

\date{Received / Accepted }

\abstract
{Ubiquitous small-scale vortices have recently been found in the lower atmosphere of the quiet Sun in state-of-the-art solar observations and in numerical simulations.}
{We investigate the characteristics and temporal evolution of a granular-scale vortex and its associated upflows through the photosphere and chromosphere of a quiet Sun internetwork region.}
{We analyzed high spatial and temporal resolution ground- and spaced-based observations of a quiet Sun region. The observations consist of high-cadence time series of wideband and narrowband images of both H$\alpha$ 6563$\,${\AA} and Ca II 8542$\,${\AA} lines obtained with the CRisp Imaging SpectroPolarimeter (CRISP) instrument at the Swedish 1-m Solar Telescope (SST), as well as ultraviolet imaging and spectral data simultaneously obtained by the Interface Region Imaging Spectrograph (IRIS).}
{A small-scale vortex is observed for the first time simultaneously in H$\alpha$, Ca II 8542$\,${\AA}, and Mg II k lines. During the evolution of the vortex, H$\alpha$ narrowband images at $-$0.77$\,${\AA} and Ca II 8542$\,${\AA} narrowband images at $-$0.5$\,${\AA}, and their corresponding Doppler Signal maps, clearly show consecutive high-speed upflow events in the vortex region. These high-speed upflows with a size of 0.5--1$\,$Mm appear in the shape of spiral arms and exhibit two distinctive apparent motions in the plane of sky for a few minutes: (1) a swirling motion with an average speed of 13$\,$km/s and (2) an expanding motion at a rate of 4--6$\,$km/s. Furthermore, the spectral analysis of Mg II k and Mg II subordinate lines in the vortex region indicates an upward velocity of up to $\sim$8$\,$km/s along with a higher temperature compared to the nearby quiet Sun chromosphere.}
{The consecutive small-scale vortex events can heat the upper chromosphere by driving continuous high-speed upflows through the lower atmosphere.}


\titlerunning{Observations of a small-scale vortex in the quiet Sun}
\authorrunning{S.-H. Park et al.}

\maketitle

\section{Introduction}
\label{sec1} Granular-scale vortical flows in quiet Sun regions are thought to be generated mainly by turbulent convection in subsurface layers of the Sun and its interaction with the solar atmosphere \citep[e.g.,][]{Stein:2000,Kitiashvili:2012a}. These small-scale vortical flows have caught significant attention in relation to the formation of magnetic or nonmagnetic vortex tubes \citep[e.g.,][]{Moll:2011,Shelyag:2011,Kitiashvili:2012b}, the formation of acoustic/Alfv\'{e}n/shock waves \citep[e.g.,][]{Fedun:2011, Kitiashvili:2011}, and the heating of the upper solar atmosphere \citep[e.g.,][]{Sturrock:1981,Zirker:1993}.

High-resolution observations of quiet Sun regions have allowed close investigation of small-scale vortex-like structures or swirling flows in the lower atmosphere. \citet{Brandt:1988} found a vortical motion of granules that persisted for $\sim$1.5 hours. \citet{Bonet:2008} detected small-scale ($<\,$0.5$\,$Mm) vortical motions in the moderately magnetized (network) regions by tracing proper motions of bright points that are engulfed by intergranular downdrafts. Likewise, \citet{Manso:2011} reported very tiny ($<\,$0.4$\,$Mm) swirl motions of internetwork magnetic elements in intergranular lanes. In addition, local correlation tracking techniques have been applied to high-cadence, high-resolution observations of the photosphere of quiet Sun internetwork regions to examine small-scale flow motions and their statistical properties \citep[e.g.,][]{Bonet:2010,Vargas:2011}. As a result, swirling motions have been found in regions of converging horizontal flows with an occurrence rate on the order of 10$^{-3}\,$Mm$^{-2}\,$min$^{-1}$. Using Ca II 8542$\,${\AA} imaging spectroscopy observations, \citet{Wedemeyer:2009} showed the presence of small-scale chromospheric swirls with an average size of $\sim$1.5$\,$Mm inside a coronal hole, consisting of dark and bright rotating patches in the form of arcs, spiral arms, rings, or ring fragments. It was also found that the chromospheric swirls exhibit upflows of 2--7$\,$km/s, indicating fast upflows, while small groups of photospheric bright points were located in intergranular lanes underneath the chromospheric swirls.

Three-dimensional (3D) radiative magnetohydrodynamics (MHD) simulations revealed that small-scale eruptions in the solar atmosphere can be driven by magnetized vortex tubes, also known as magnetic tornadoes \citep[e.g.,][]{Wedemeyer:2012,Kitiashvili:2013,Wedemeyer:2014}. In particular, \citet{Kitiashvili:2013} showed that small-scale plasma eruptions occur in swirling vortex tubes generated by the Sun's turbulent convection in subsurface layers. They also found a complicated structure and dynamics of the small-scale eruptions in which the flows are predominantly downward in the vortex cores and upward around the periphery of the vortex cores. The shape of the eruptions in the vortex tube resembles swirls or tornadoes and the eruptions are quasiperiodic with a characteristic period of 2--5 minutes. However, because of the small number of vortex events reported from observations, the detailed characteristics of small-scale vortical flow eruptions are still not well understood. 

In this study, we investigate granular-scale vortices in a quiet Sun internetwork region observed for the first time on high-resolution, high-cadence filtergrams (in H$\alpha$ and Ca II 8542$\,${\AA}) and spectrograms (in Mg II) obtained simultaneously with state-of-the-art ground-based and space solar instruments (CRISP and IRIS, respectively). Our aim is to provide a better understanding of the complicated dynamics of small-scale vortical flows and their temporal evolution in the photosphere and chromosphere of quiet Sun regions.

\section{Observations and data analysis}
\label{sec2} Coordinated observations with the Swedish 1-m Solar Telescope \citep[SST;][]{Scharmer:2003a} and the Interface Region Imaging Spectrograph \citep[IRIS;][]{DePontieu:2014a} were carried out on June 7, 2014 between 07:32 UT -- 08:21 UT to obtain a multiwavelength data set of a quiet Sun region with high spatial and temporal resolution.

\begin{figure}
\centering
\includegraphics[width=9cm]{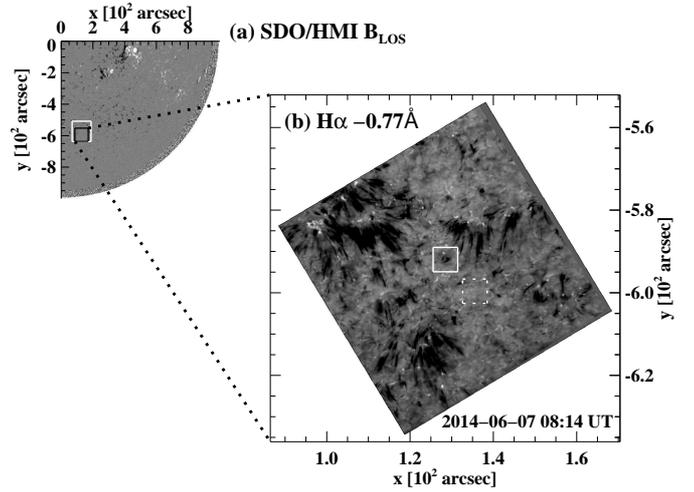}
\caption{(a) HMI line-of-sight magnetogram with the FOVs of the CRISP and IRIS observations overplotted in black and white rectangles, respectively. (b) A sample of the cotemporal CRISP FOV H$\alpha-$0.77$\,${\AA} image. The overplotted white solid and dashed rectangles in panel b indicate the ROI (i.e., the FOV of Figs.~\ref{f2}--~\ref{f4}{a} and column 1--3 of Fig.~\ref{f6}) and a neighboring typical quiet Sun region considered in determining the average H$\alpha$ line reference profile, respectively.} \label{f1}
\end{figure}

In the present study, we use high-cadence (i.e., 4\,seconds) time series of wideband and narrowband images of both H$\alpha$ and Ca II 8542$\,${\AA} lines taken with the CRisp Imaging SpectroPolarimeter \citep[CRISP;][]{Scharmer:2008} at SST. The CRISP samples (1) the H$\alpha$ line with a wideband filter of 4.9$\,${\AA}; (2) the H$\alpha$ line at seven wavelength positions (i.e., the line center, H$\alpha\pm$0.26$\,${\AA}, H$\alpha\pm$0.77$\,${\AA}, and H$\alpha\pm$1.03$\,${\AA}) with a narrowband filter of 0.066$\,${\AA}; (3) the Ca II 8542$\,${\AA} line using a wideband filter of 9.3$\,${\AA}; and (4) the Ca II 8542$\,${\AA} line at seven wavelength positions (i.e., the line center, Ca II$\pm$0.055$\,${\AA}, Ca II$\pm$0.11$\,${\AA}, and Ca II$\pm$0.495$\,${\AA}) using a narrowband filter of 0.111$\,${\AA}. The pixel sizes of the images are 0.059$\arcsec$ and 0.0576$\arcsec$ for the H$\alpha$ and Ca II 8542$\,${\AA} lines, respectively. The image quality of the time series benefited from the SST adaptive optics system \citep{Scharmer:2003b} and image restoration technique Multi-Object Multi-Frame Blind Deconvolution \citep[MOMFBD;][]{Noort:2005}. We use early versions of the CRISP data reduction pipeline \citep[CRISPRED;][]{Cruz:2015} for preparation of the H$\alpha$ and Ca II 8542$\,${\AA} spectral datacubes. The field of view (FOV) of the CRISP observations covers a quiet Sun area of about 60$\arcsec\times\,$60$\arcsec$ and the images are rotated around the center, corresponding to (x,$\,$y) $=$ ($-$128$\arcsec$,$\,-$594$\arcsec$) to correct for solar tilt. Figure~\ref{f1}{a} shows the position of the CRISP full FOV (indicated with a black rectangle) on a co-aligned and cotemporal line-of-sight magnetogram observed by the Helioseismic and Magnetic Imager \citep[HMI;][]{Scherrer:2012} aboard the Solar Dynamics Observatory \citep[SDO;][]{Pesnell:2012}. 

We also investigate simultaneous ultraviolet (UV) imaging and spectroscopic data obtained with IRIS. The imaging system of IRIS provides slit-jaw images (SJIs) in four different channels (1330$\,${\AA} and 1400$\,${\AA} with a 40$\,${\AA} bandpass and 2796$\,${\AA} and 2832$\,${\AA} with a 4$\,${\AA} bandpass): however, during these coordinated observations, only the 1400$\,${\AA} SJIs were obtained with a 17\,sec cadence in order to increase the S/N ratio. In addition, the IRIS slit-based spectrograph observes the chromosphere and transition region of the Sun: (1) in two far UV wavelength ranges (1331.6$\,$--$\,$1358.4$\,${\AA} at 12.98 m{\AA}/pixel and 1380.6$\,$--$\,$1406.8$\,${\AA} at 12.72 m{\AA}/pixel) including C II, O I and Si IV lines, and (2) in one near UV range (2782.6$\,$--$\,$2833.9$\,${\AA} at 25.46 m{\AA}/pixel) containing Mg II k $\&$ h lines. The size of the slit is 0.33$\arcsec\times\,$175$\arcsec$. The IRIS far/near UV spectra were obtained in a sit-and-stare mode with a cadence of 17\,sec and an exposure time of 15\,sec. Exposure times of  4--8\,sec are typically employed for IRIS observations of the quiet Sun \citep[e.g.,][]{DePontieu:2014b,Voort:2015}. We only use the Mg II k 2796$\,${\AA} line (logT $=$ 4$\,$K) spectra in this study.

The SST/CRISP and IRIS data were co-aligned by (1) correcting the drift in the IRIS pointing using cross-correlation between successive IRIS 1400$\,${\AA} SJIs, (2) degrading the spatial resolution of the CRISP H$\alpha$ and Ca II 8542$\,${\AA} images to the IRIS spatial pixel size of 0.166$\arcsec$, and (3) comparing the Ca II$-$0.5$\,${\AA} far blue wing images with the H$\alpha-$1.03$\,${\AA} far blue wing images and the 1400$\,${\AA} SJIs, respectively, for similar network and internetwork features. This co-alignment process proved to be accurate down to the level of the IRIS pixel scale. The co-alignment between SST/CRISP and IRIS was double-checked using full-disk 1600$\,${\AA} images from the Atmospheric Imaging Assembly \citep[AIA;][]{Lemen:2012} on board SDO, as a common reference.

\begin{figure*}
\centering
\includegraphics[width=18cm,clip]{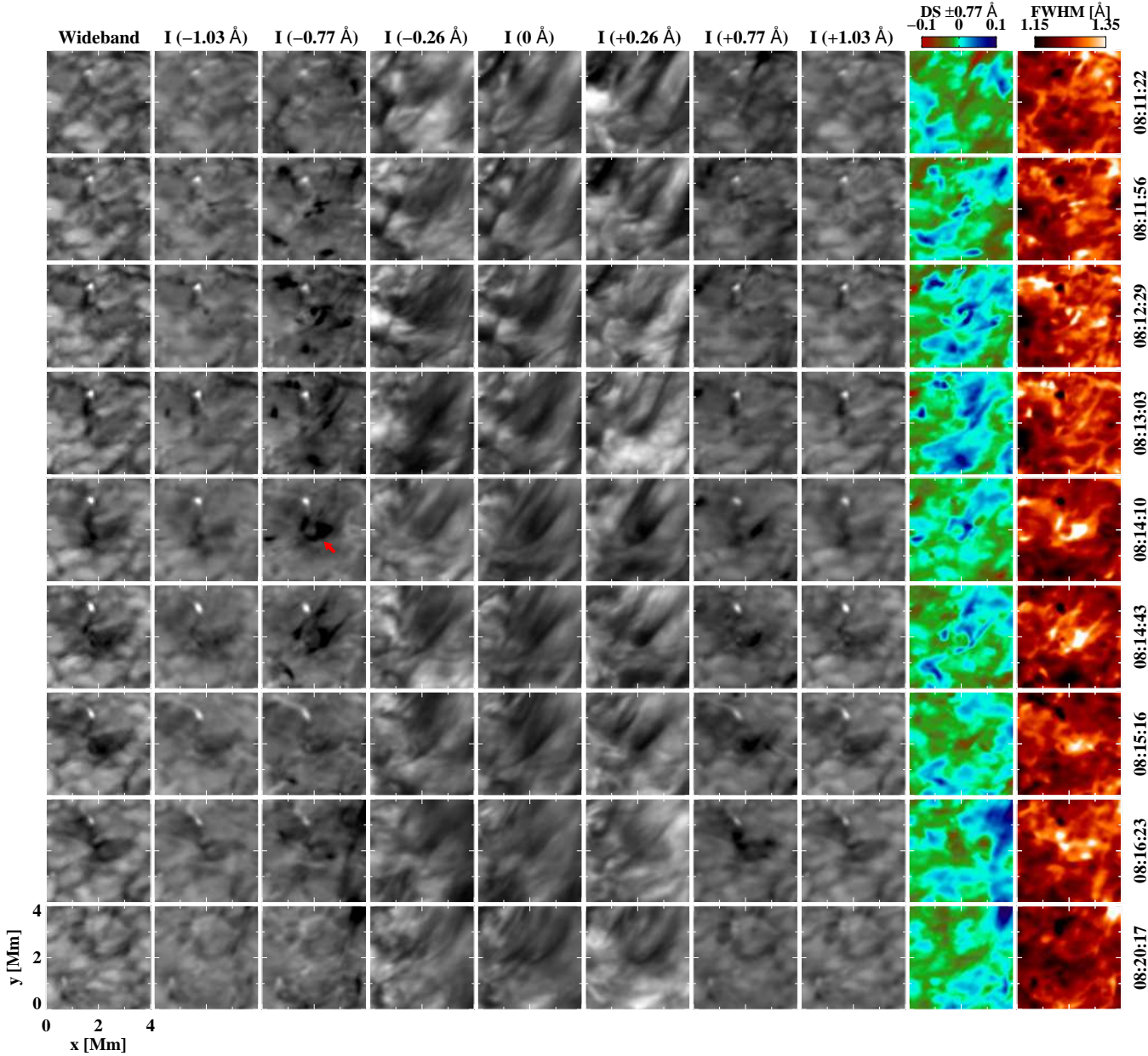}
\caption{Sequence of the H$\alpha$ wideband (column 1) and narrowband (columns 2--8: $-$1.03$\,${\AA}, $-$0.77$\,${\AA}, $-$0.26$\,${\AA}, the H$\alpha$ line center, $+$0.26$\,${\AA}, $+$0.77$\,${\AA} and $+$1.03$\,${\AA}) images of the ROI, which is indicated with a solid box in Fig.~\ref{f1}{b}, including the respective maps of DS at $\pm$0.77$\,${\AA} (column 9) and FWHM (column 10). Respective times of observation are shown along the right ordinate. This figure is accompanied by an on-line movie. The rightmost column for the FWHM, which is present in Fig.~\ref{f2}, is not included in the movie.} \label{f2}
\end{figure*}

In our analysis, we calculated the Doppler Signal (DS) from the CRISP H$\alpha$ and Ca II profiles. DS provides a qualitative picture of upward (positive DS) and downward (negative DS) moving material \citep{Tsiropoula:2000}, and is defined as follows:
\begin{equation}
\label{sec2:eq1} DS =
\frac{I(+\Delta\lambda)-I(-\Delta\lambda)}{I(+\Delta\lambda)+I(-\Delta\lambda)},
\end{equation}
where $I(\pm\Delta\lambda)$ is the intensity of the H$\alpha$ or Ca II off-band images at $\pm\Delta\lambda$ from the line center. A zero reference of DS is defined as the mean DS value of a neighboring quiet Sun region indicated with the white dashed rectangle in Fig.~\ref{f1}{b} to correct for the projection effect. We also determined the full width at half maximum (FWHM) of the H$\alpha$ profile in each pixel of the area under study by measuring the wavelength separation between two wavelength positions on either side of the profile, where the spectrum intensity $I$ equals 0.5$\times(I_{C}+I_{min})$. The intensities $I_{C}$ and $I_{min}$ are the continuum and minimum intensities of the profile. We determine $I_{C}$ by fitting the average H$\alpha$ profile in the neighboring quiet sun region (indicated with the dashed rectangle in Fig.~\ref{f1}{b}) to a reference full H$\alpha$ profile \citep{David:1961}, which covers up to $\pm$30$\,${\AA} from the line center. 

\section{Results}
\label{sec3} 
\subsection{Small-scale vortex in H$\alpha$ and Ca II 8542$\,${\AA}}
\label{sec3_1} Figure~\ref{f1}{b} shows the CRISP full FOV H$\alpha-$0.77$\,${\AA} image at 08:14:26 UT. The observed area is covered predominantly by several dark mottles and, as typically seen in quiet Sun observations, consists of bright points in network regions and of granules and intergranular jets, which appear as cloud absorption features, in internetwork regions. In a small area of the FOV, we detected an interesting small-scale vortex located in an internetwork region, but well separated from neighboring mottles. The white solid rectangle in Fig.~\ref{f1}{b} indicates the region-of-interest (ROI) of 4$\,$Mm$\times\,$4$\,$Mm that contains the vortex and that we analyze in detail in what follows.

\begin{figure*}
\centering
\includegraphics[width=18cm,clip]{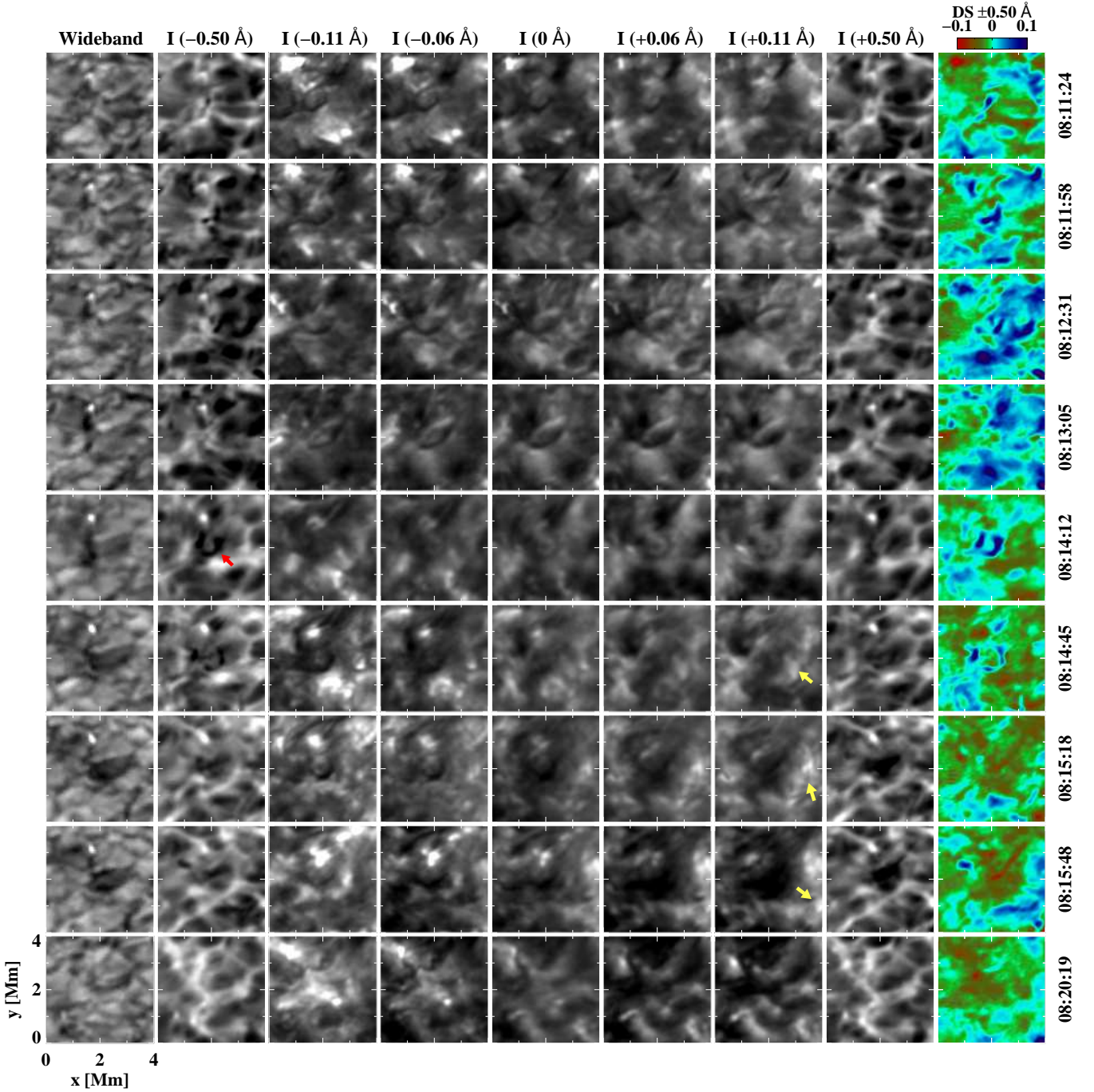}
\caption{Sequence of the Ca II 8542$\,${\AA} wideband (column 1) and narrowband (columns 2--8: $-$0.5$\,${\AA}, $-$0.11$\,${\AA}, $-$0.06$\,${\AA}, the Ca II 8542$\,${\AA} line center, $+$0.06$\,${\AA}, $+$0.11$\,${\AA} and $+$0.5$\,${\AA}) images of the ROI, including DS maps at Ca II$\pm$0.5$\,${\AA} (column 9). Respective times of observation are shown along the right ordinate. Some arc-shaped propagating bright features are indicated with arrows in column 7. This figure is accompanied by an on-line movie.} \label{f3}
\end{figure*}

Figure~\ref{f2} shows the sequence of the H$\alpha$ wideband (column 1) and narrowband images (columns 2--8) of the ROI at some particular times during a period of $\sim$10\,min from 08:11 UT to 08:20 UT. In general, vortex-like or ring-like features are observed at the center of the wideband and narrowband images. The vortex is clearly seen in the H$\alpha-$0.77$\,${\AA} images in absorption (see the on-line movie) consisting of spiral arms (shown, e.g., with red arrow at 08:14:10 UT), while in the H$\alpha+$0.77$\,${\AA} images, localized absorption features appear in and around the vortex core (e.g., at 08:15:16). At the H$\alpha$ line center and H$\alpha\pm$0.26$\,${\AA} there is still an imprint of the vortex, although this imprint is faint and with reduced contrast, but reminiscent of what is seen in the H$\alpha-$0.77$\,${\AA} images. The H$\alpha$ wideband images are characterized by one very bright point, granules and intergranular lanes, while the vortex appears as a very dark large granule. The wideband images are very similar to those corresponding to H$\alpha\pm$1.03$\,${\AA} because the filter used for the wideband only covers the wavelength range of $-$2.5$\,${\AA} to $+$2.5$\,${\AA} from the H$\alpha$ line center.

In Figure~\ref{f2} the respective maps of DS at H$\alpha\pm$0.77$\,${\AA} (column 9) and FWHM of the H$\alpha$ line profiles (column 10) in the area of the ROI are also shown. The DS maps indicate that the spiral arms of the vortex shown at H$\alpha-$0.77$\,${\AA} consist of high-speed upflowing material, while the localized absorption features of the vortex core seen at H$\alpha+$0.77$\,${\AA} consist mainly of downflowing material. Moreover, the FWHM maps (last column of Fig.~\ref{f2}) show that the H$\alpha$ line profiles in the vortex region have relatively larger FWHM values compared to those in the surrounding region, which could be attributed to a higher microturbulence and/or a higher temperature of the vortex plasma.

\begin{figure}
\centering
\includegraphics[width=8.8cm]{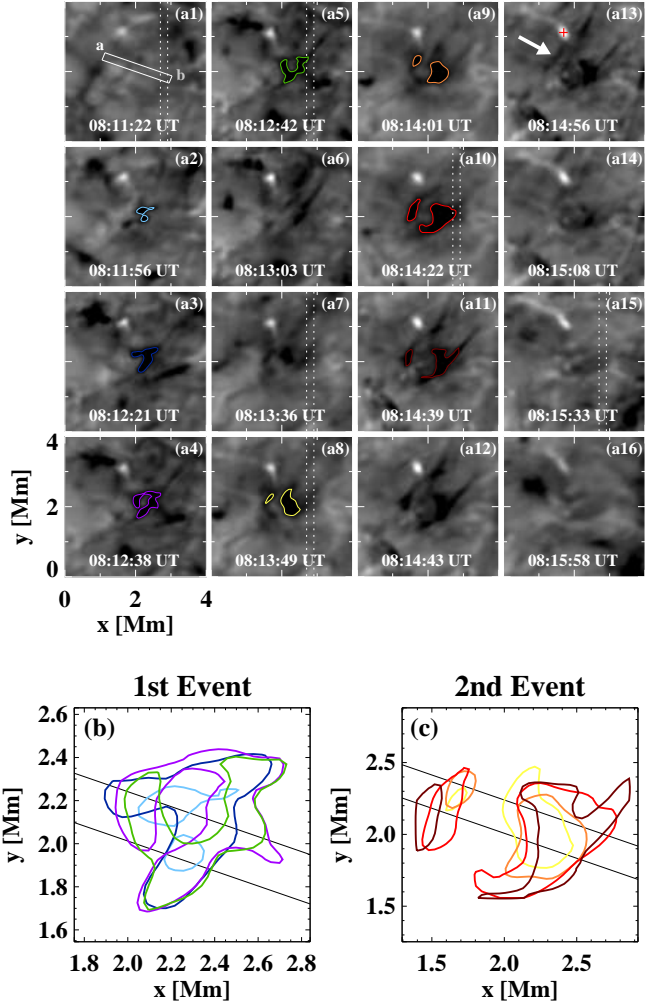}
\caption{Temporal evolution of the two consecutive upflow events. The sequence of the H$\alpha-$0.77$\,${\AA} images of the ROI during the two upflow events is shown in panels a1--16. The 2$\,$Mm slit (a--b) in panel a1 indicates the cut location used for the space-time slice image shown in Fig.~\ref{f5}. The arrow in panel a13 denotes a top left spiral arm, which encounters a neighboring bright point indicated with a red cross in the same panel. The locations of the cotemporal IRIS slit are indicated with two vertical dotted lines on several panels, which are used for the investigation of the IRIS Mg II k spectrum in Figs.~\ref{f6} and~\ref{f7}. The outlines of the vortical structures in some of panels a1--a16 during the first and second events are overplotted with different colored contours in panels b and c, respectively.} \label{f4}
\end{figure}

We also examine the simultaneous data set of the CRISP Ca II 8542$\,${\AA} imaging spectroscopy observations. Figure~\ref{f3} shows the sequence of the Ca II 8542$\,${\AA} wideband (column 1) and narrowband images (columns 2--8) of the ROI along with the respective DS maps at Ca II$\pm$0.5$\,${\AA} (column 9), at similar times as in Fig.~\ref{f2}. Similar features of the vortex as in the H$\alpha$ images are observed at the center of the ROI. In particular, the upflow features of the vortex spiral arms seen at H$\alpha-$0.77$\,${\AA} (column 3 of Fig.~\ref{f2}) are clearly seen from $\sim$08:12 to $\sim$08:14 UT at the Ca II$-$0.5$\,${\AA} images (indicated with a red arrow at the corresponding panel at 08:14:12 UT; see also the on-line movie) and the respective DS maps at Ca II$\pm$0.5$\,${\AA}. The Ca II wideband appearance is similar to the H$\alpha$ windeband appearance. At the Ca II line center and both observed red and blue wings close to the line center, the vortex appears like dark patches in the form of elliptical rings or ring fragments, which are similar to the small-scale chromospheric swirls reported by \citet{Wedemeyer:2009}. We also find arc-shaped propagating bright features (denoted by the arrows in column 7 of Fig.~\ref{f3}) at almost all seven wavelengths (more pronounced at the red wings), which are adjacent to the vortex. Similar bright features are observed in the IRIS 1400$\,${\AA} SJIs, which propagate with a speed of $\sim$7--13$\,$km/s. We will investigate these vortex-driven propagating features in a follow-up study.

A detailed examination of the vortex-associated upflows and their temporal evolution is now carried out with the H$\alpha-$0.77$\,${\AA} images. During the evolution of the vortex, we detect two consecutive upflow events in similar locations around the vortex core. Each of them is composed of a pair of two spiral arms that first appear very tiny, then continue to expand in size and finally fade away. Figure~\ref{f4}{a} shows the sequence of the H$\alpha-$0.77$\,${\AA} images of the ROI during the period of the upflow events. The first upflow event starts at 08:11:56 UT with two tiny elongated features that appear in the intersection (i.e., intergranular lanes) of neighboring granules parallel to each other. This shape resembles to the one called ``Type II'' by \citet{Wedemeyer:2013} in their Fig.~\ref{f3}, where they also note, ``Yet to be found in observations''. Then these two elongated features keep on developing into a vortex that grows in size and goes through swirling and expanding motions. Right after the first upflow event completely disappears at 08:13:03 UT, the second upflow event starts in the immediate vicinity of the first event. It develops into two separated spiral arms experiencing a similar pattern of evolution until 08:14:56 UT, as seen in the first upflow event. As one of the two spiral arms (shown with an arrow in panel a13 of Fig.~\ref{f4}) encounters a neighboring bright point (indicated with a red cross in the same panel), it fades away rapidly while the bright point gets stretched or extended along the intergranular lane in the direction where the spiral arm propagates. The second upflow event finishes at 08:15:16 UT and then a ring-like feature remains in the vortex core for $\sim$2 minutes, shown as downflowing material in the DS map at H$\alpha\pm$0.77$\,${\AA}. The maximum apparent diameters (lifetimes) of the first and second vortex events are 0.6$\,$Mm (1 minute) and 1.4$\,$Mm (2 minutes), respectively. In panels b and c of Fig.~\ref{f4}, we overplot the outlines of the two spiral arms, which are traced and denoted with different colored contours in Fig.~\ref{f4}{a} during the first and second upflow events, respectively, so as to ascertain their two characteristic apparent motions: i.e., swirling and expanding motions. The swirling motion around the vortex center can be seen in the top left part of the two spiral arms during the first event (see Fig.~\ref{f4}{b}), and this motion is traced using the slit (a--b) shown in panel a1 of Fig.~\ref{f4} (see more detailed description below). The expanding motion, in which the two spiral arms move away from each other, is also well detected during both events, especially along the slit direction.

\begin{figure}
\centering
\includegraphics[width=8.5cm]{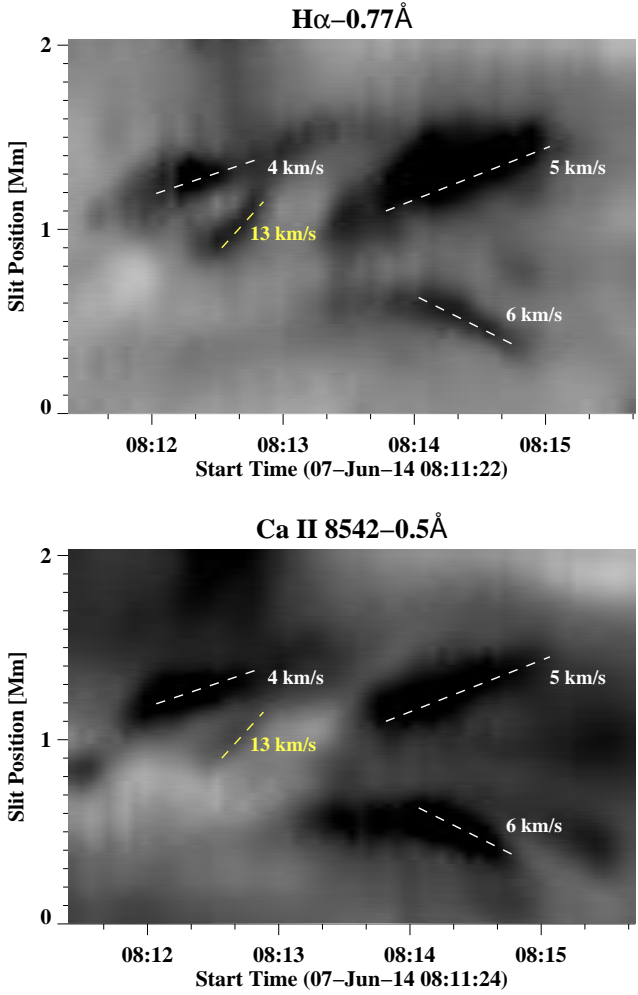}
\caption{Space-time slice image of (a) H$\alpha-$0.77$\,${\AA} intensity and (b) Ca II 8542$-$0.5$\,${\AA} intensity along the slit a--b shown in the first panel of Fig.~\ref{f4}. The yellow and white dashed lines indicate the swirling and expanding motions, respectively, which are detected along the slit during the two consecutive upflow events.}
\label{f5}
\end{figure}

\begin{figure}
\centering
\includegraphics[width=8.8cm]{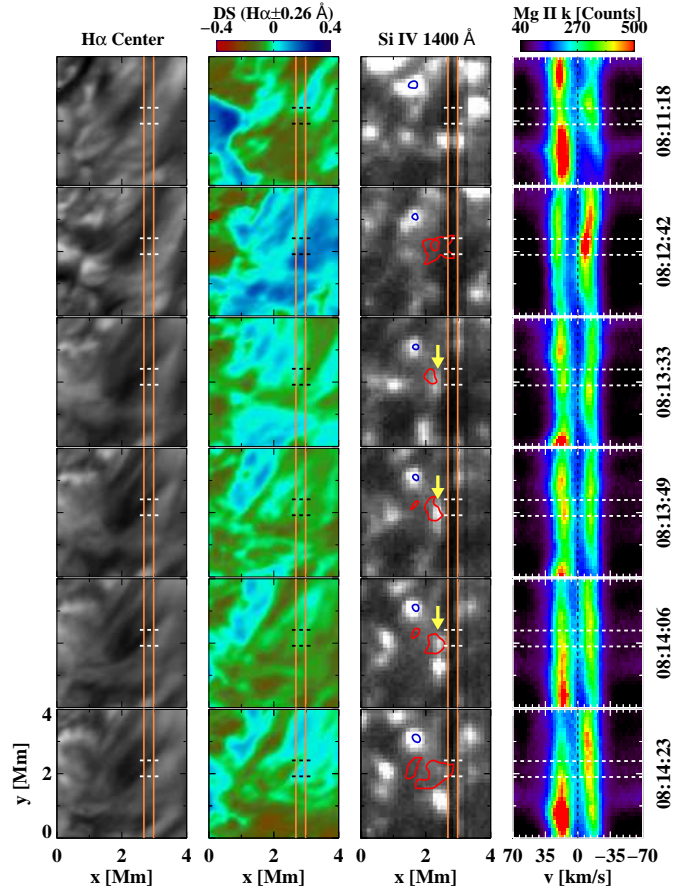}
\caption{Simultaneous H$\alpha$ and UV observations of the vortex region. The H$\alpha$ line center images (column 1), DS maps at H$\alpha\pm$0.26$\,${\AA} (column 2), IRIS Si IV 1400$\,${\AA} SJIs (column 3) of the ROI, as well as the respective IRIS Mg II k 2796$\,${\AA} spectra (column 4), are shown at certain times before and during the two upflow events. The red and blue contours on the SJIs represent the two spiral arms of the vortex and their neighboring bright point seen in the H$\alpha-$0.77$\,${\AA} images, respectively. The IRIS slit is indicated with two vertical lines in each panel in columns 1--3. A subregion on the slit is shown with a dashed strip in each panel, which is used for determining the mean line profiles of Mg II k 2796$\,${\AA} and subordinate Mg II 2798.8$\,${\AA} lines in Fig.~\ref{f7}. A positive (negative) Doppler velocity corresponds to an upflow (downflow).}
\label{f6}
\end{figure}

Applying space-time slice image analysis, we quantitatively investigate the dynamics of the two consecutive upflow events. Figure~\ref{f5} shows the space-time slice image of (a) H$\alpha-$0.77$\,${\AA} intensity and (b) Ca II 8542$-$0.5$\,${\AA} intensity along the slit a--b indicated in panel a1 of Fig.~\ref{f4}. The two upflow events are seen in this space-time slice image to begin at $\sim$08:12 UT and $\sim$08:14 UT, respectively. The slit has a length of 2$\,$Mm passing through the center of each pair of the two vortical arms during the two upflow events. Also, during the investigation time, the slit is located in the direction parallel to the vortex's expanding motion except in one case during the first upflow event in which the slit is along the direction of the swirling motion of the top left spiral arm as shown in Fig.~\ref{f4}{b}. As a result, we can examine the details of the two characteristic flow motions seen in the blue wings of both H$\alpha$ and Ca II 8542$\,${\AA} lines. From inspecting the space-time slice images, we see that the flow motions detected in the H$\alpha-$0.77$\,${\AA} space-time plot are almost identical to those in the Ca II 8542$-$0.5$\,${\AA} space-time plot without any substantial time lag between them (i.e., almost in phase). The swirling motion of the first upflow event, detected from the counterclockwise vortical flow along the slit direction, is indicated with a yellow dashed line in each panel of Fig.~\ref{f5}. It is sustained during a period of $\sim$40 seconds with an average speed of $\sim$13$\,$km/s. On the other hand, the expanding motion of each pair of the two spiral arms is measured during both uplfow events. In particular, it is more clearly seen during the second event over a span of $\sim$2 minutes. The average speed of the expanding motion is estimated to be $\sim$4--6$\,$km/s. 

\subsection{Analysis of IRIS SJIs and spectra}
\label{sec3_2} We also investigate the IRIS imaging and spectroscopic observations of the vortex region. Figure~\ref{f6} shows the narrowband images of the H$\alpha$ line center (column 1), the DS maps at H$\alpha\pm$0.26$\,${\AA} (column 2), and the IRIS 1400$\,${\AA} SJIs (column 3) of the ROI, as well as the respective IRIS Mg II k 2796$\,${\AA} spectra (column 4), at some particular times before and during the two upflow events. The contours of spiral arms of the vortex and its neighboring bright point seen at the H$\alpha-$0.77$\,${\AA} images are overplotted on the SJIs with red and blue contours, respectively. The bright point appears in both H$\alpha-$0.77$\,${\AA} images and SJIs at exactly the same location, indicating that the H$\alpha$ images are well aligned with the SJIs. The position of the IRIS slit is shown with the two vertical lines in each panel of columns 1--3. As shown in the SJIs, the edge of the bottom right spiral arm of each upflow event is sometimes located under the IRIS slit which has a width of 0.24$\,$Mm (e.g., at 08:12:42 UT and 08:14:23 UT for the first and second event, respectively). We checked all the available IRIS spectra (including chromospheric lines, such as Mg II k 2796$\,${\AA} and Mg II h 2803$\,${\AA} and far UV lines, such as C II 1334/1335$\,${\AA}, O I 1356$\,${\AA} and Si IV 1394/1403$\,${\AA}) during the period of the upflow events, but we only find a meaningful signal at the vortex region in the Mg II k $\&$ h spectra. Given that the exposure time used for the IRIS observations is rather high, we conclude that no meaningful signal at the vortex region in the FUV lines is not due to noise, but most probably because the upflow events do not reach the transition region. The IRIS Mg II k $\&$ h spectrum data show very similar spectral profiles so that we only present the Mg II k spectra. As shown in column 4 of Fig.~\ref{f6}, the Mg II k spectra can be generally characterized by a central absorption core (k$_{3}$) surrounded by two emission peaks (k$_{2V}$ and k$_{2R}$ for the red and blue sides of the peaks, respectively). The rest wavelength is defined as the mean wavelength of the central absorption cores observed in the full FOV of the IRIS spectrograph except the vortex region. We use the convention that a positive (negative) Doppler shift corresponds to a blueshift (redshift) and an upflow (downflow) in order to be consistent with the definition of the DS in Eq. (1).

In the SJIs, we find that some elongated bright features (shown with the yellow arrows in column 3) appear in the vicinity of the bottom right part of the two spiral arms during the second upflow event, even though they do not clearly show a vortex-like structure. Through direct inspection of the Mg II k spectra (column 4), we see that as the slit crosses the vortex edges (column 3), the line core k$_{3}$ is shifted toward the blue (i.e., upflow) from the rest wavelength at particular times, e.g., at 08:12:42 UT and 08:14:23 UT. We also see that the k$_{2R}$ peak is enhanced at the vortex edges, especially during the first upflow event. It is important to assure that the Mg II k spectra observed in the vortex edges reflect the properties of the vortex seen in the H$\alpha$ and Ca II 8542$\,${\AA} lines and not the properties of an overlying structure. Inspecting the H$\alpha$ line center images, we conclude that there are no any large-scale overlying features (such as mottles) at the vortex edges during the period of the two upflow events. Instead, two large elongated vortical structures, which seem to be an imprint of the two spiral arms of the vortex at the H$\alpha-$0.77{\AA} are clearly seen at the H$\alpha$ line center. In addition, the Doppler shift of k$_{3}$ in the Mg II k spectra is qualitatively similar to the DS at H$\alpha\pm$0.26$\,${\AA} not only at the vortex edges, but also at the rest of the region along the slit. This suggests that the Mg II k spectra reflect the spectral properties of the plasma associated with the vortex. 

\begin{figure}
\centering
\includegraphics[width=8.8cm]{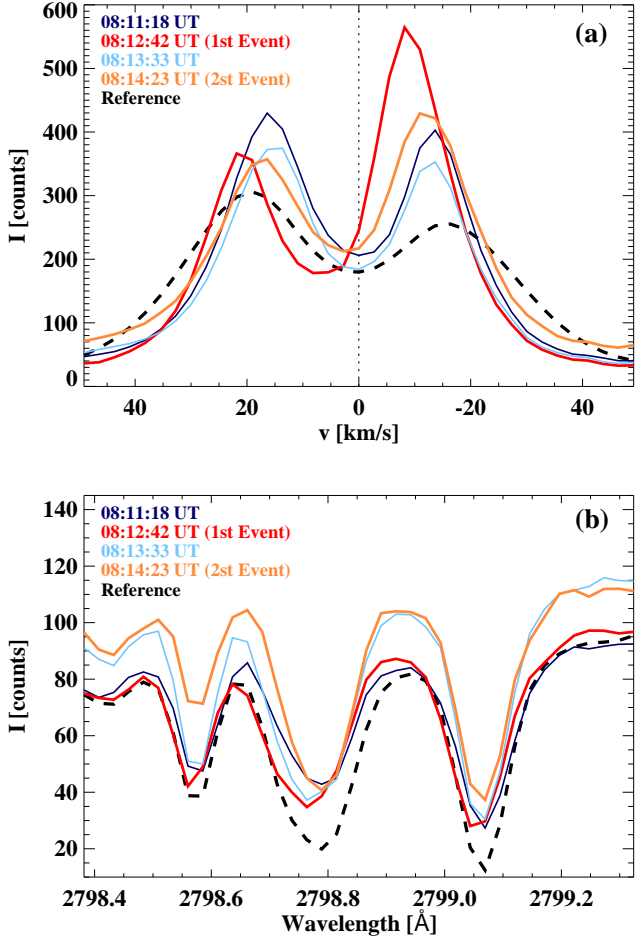}
\caption{Mean line profiles of (a) Mg II k 2796$\,${\AA} and (b) subordinate Mg II 2798.8$\,${\AA} lines in a subregion of the ROI, indicated with the strip in Fig.~\ref{f6}, at some specific times including the observations of the vortex edges by the IRIS slit (i.e., 08:12:42 UT and 08:14:23 UT for the first and second upflow event, respectively). The mean line profile (black dashed line) in the full FOV of the IRIS spectrograph excluding the vortex region is also shown at each panel as a reference.}
\label{f7}
\end{figure}

We further analyze some spectral properties of the vortex plasma from the observed Mg II k line profiles, applying the diagnostic technique of \citet{Leenaarts:2013} for the Mg II k line in a dynamic 3D radiative MHD model of the solar atmosphere. Figure~\ref{f7} shows the mean line profiles of (a) Mg II k 2796$\,${\AA} and (b) subordinate Mg II 2798.8$\,${\AA} lines in the subregion, indicated with a strip in Fig.~\ref{f6}, at the same times as in Fig.~\ref{f6}. The reference line profile (black dashed line) averaged in the full FOV of the IRIS spectrograph, excluding the vortex region, is overplotted in each panel. We first investigate the Doppler shift of the k$_{3}$ line center. In \citet{Leenaarts:2013}, it was found that this Doppler shift shows a strong correlation (i.e., Pearson correlation coefficient of 0.99) with the vertical velocity at optical depth unity, which is typically located less than 200$\,$km below the transition region. We find that the Doppler shifts of k$_{3}$ measured from the line profiles of the vortex edges are 8$\,$km/s and 3$\,$km/s for the first and second upflow events, respectively. The speed of these upflows is similar to that of the apparent vortex motions. There is another way to check whether the vortex plasma seen in the Mg II k line indeed moves upward. \citet{Leenaarts:2013} found that the average velocity in the upper chromosphere correlates strongly with the intensity ratio $R_{k}$ of the blue (k$_{2V}$) and red (k$_{2R}$) peaks defined as follows:
\begin{equation}
\label{sec3:eq2} R_{k} =
\frac{I_{\mathrm{k}_{2V}}-I_{\mathrm{k}_{2R}}}{I_{\mathrm{k}_{2V}}+I_{\mathrm{k}_{2R}}},
\end{equation}
where a positive (negative) peak ratio corresponds to a downflow (upflow). The line profiles of the vortex edges show that the red k$_{2R}$ peak is higher than the blue k$_{2V}$ peak with $R_{k}$ of $-$0.2 to $-$0.1 for the first and second upflow events, respectively. \citet{Leenaarts:2013} also showed that the intensity of both k$_{2V}$ and k$_{2R}$ peaks have a good correlation with the temperature at the height of optical depth unity so that the peaks can be exploited as a temperature diagnostic. The blue and red peaks of the vortex edge profiles are higher than those of the average reference profile, where the difference is almost twice in the case of the red peak. This suggests that the vortex plasma has a higher temperature compared to the background plasma temperature of the quiet Sun. Moreover, we check the Mg II subordinate lines at 2798.8$\,${\AA}, which can be used for a temperature diagnostic of the chromosphere \citep[refer to][]{Pereira:2015}. The subordinate line at 2798.8$\,${\AA} is in absorption and during the first event its intensity is very close to that of the reference line, unlike the intensity profile of the Mg II k line (Fig.~\ref{f7}{a} and {b}, red line), which is highly increased. During the second upflow event, however, the subordinate line is strongly enhanced relative to the reference profile similar to the intensity profile of the Mg II k line (Fig.~\ref{f7}{a} and {b}, orange line). Given that the subordinate line is formed lower than the Mg II k line, the intensity increase in the Mg II k line during the first event could be caused by a temperature increase at the chromosphere, while the intensity enhancements observed in both lines during the second event could be caused by a temperature increase throughout a wide range of the atmosphere \citep{Pereira:2015}. This is consistent with the fact that the enhancement of the FWHM of the H$\alpha$ line profile (corresponding to a layer lower than the chromosphere) in the vortex region is more evident during the second event, as shown in the last column of Fig.~\ref{f2}, which we attribute to a higher microturbulence and/or a higher temperature of the vortex plasma.

\section{Summary and conclusions}
\label{sec4} We investigated the dynamics and evolution of consecutive upflow events in a small-scale quiet Sun vortex with simultaneous, high-resolution and high-cadence observations of a quiet Sun region obtained by SST/CRISP and IRIS. The important characteristics of the observed upflow events can be summarized as follows:
\begin{enumerate}  
\item Two consecutive upflow events occur in a small-scale vortex during a period of $\sim$5 minutes and are observed for the first time in both H$\alpha$ and Ca II lines;
\item Each of them consists of high-speed upflow features in the form of a pair of two spiral arms with a spatial scale of $\sim$1$\,$Mm (sometimes even less than 1$\,$Mm);
\item Two different types of apparent motions of the upflow features are detected: i.e., swirling and expanding motions with an average speed of $\sim$13$\,$km/s and $\sim$5$\,$km/s, respectively;
\item The spectral analysis of Mg II k and Mg II subordinate lines in the vortex region indicates that the vortex plasma at the upper chromosphere reaches an upward velocity up to $\sim$8$\,$km/s, while its temperature increases during the upflow events.
\end{enumerate}

Through this observational study on the vortex and its associated upflows, we conclude that (1) consecutive granular-scale upflows can be generated in the quiet Sun, and extend through the photosphere up to the chromosphere; (2) they can develop over a short time span of a few minutes through a series of swirling and/or expanding motions; and (3) mass and energy could be transported into the upper chromosphere by the vortex-associated, high-speed upflows. We emphasize that the vortex can be followed from the photospheric internetwork up to the chromosphere of the quiet Sun, however, no signature of the vortex was detected in IRIS spectra of transition-region lines. It is also worthwhile to mention that during the upflow events we find vortex-driven features around the vortex region in the form of propagating bright arcs in the Ca II 8542$\,${\AA} narrowband images and in the SJIs. This is an interesting observational result and will be the subject of a follow-up work.

It is not clear from this study if the small-scale vortex under investigation has any relationship to magnetic tornadoes \citep[e.g., reported in][]{Wedemeyer:2012,Wedemeyer:2014}. If so, the 3D structure and temporal evolution of the magnetic field in the vortex region would be of particular interest. Magnetic tornadoes have rotating magnetic field structures that are rooted in the top layers of the convection zone and extend throughout the atmosphere up to the low corona. Unfortunately, however, in the present case it was not possible to find any clues about the magnetic field structure and evolution as a result of the lack of simultaneous high-resolution magnetic field observations. Further observational studies including magnetic fields are certainly needed to more precisely understand the origin and evolution of small-scale vortex structures and vortex-associated upflows along with their effect on possible generation of shock waves and the subsequent heating of the solar atmosphere.

\begin{acknowledgements}
The authors would like to thank the anonymous referee for many constructive comments. The research was partly funded through the project ``SOLAR-4068'', which is implemented under the ``ARISTEIA II'' Action of the operational program ``Education and Lifelong Learning'' and is cofunded by the European Social Fund (ESF) and Greek national funds. Armagh Observatory is grant aided by the N. Ireland Department of Culture, Arts, and Leisure. The Swedish 1-m Solar Telescope is operated on the island of La Palma by the Institute for Solar Physics of Stockholm University in the Spanish Observatorio del Roque de los Muchachos of the Instituto de Astrofsica de Canarias. The authors wish to acknowledge the DJEI/DES/SFI/HEA Irish Centre for High-End Computing (ICHEC) for the provision of computing facilities and support. We also like to thank STFC PATT and Solarnet project, which is supported by the European Commission's FP7 Capacities Programme under Grant Agreement number 312495 for T$\&$S.
\end{acknowledgements}


\bibliographystyle{aa} 
\bibliography{ms_bib}

\begin{thebibliography}{34}
\expandafter\ifx\csname natexlab\endcsname\relax\def\natexlab#1{#1}\fi

\bibitem[{{Bonet} {et~al.}(2008){Bonet}, {M{\'a}rquez}, {S{\'a}nchez Almeida},
  {Cabello}, \& {Domingo}}]{Bonet:2008}
{Bonet}, J.~A., {M{\'a}rquez}, I., {S{\'a}nchez Almeida}, J., {Cabello}, I., \&
  {Domingo}, V. 2008, \apjl, 687, L131

\bibitem[{{Bonet} {et~al.}(2010){Bonet}, {M{\'a}rquez}, {S{\'a}nchez Almeida},
  {Palacios}, {Mart{\'{\i}}nez Pillet}, {Solanki}, {del Toro Iniesta},
  {Domingo}, {Berkefeld}, {Schmidt}, {Gandorfer}, {Barthol}, \&
  {Kn{\"o}lker}}]{Bonet:2010}
{Bonet}, J.~A., {M{\'a}rquez}, I., {S{\'a}nchez Almeida}, J., {et~al.} 2010,
  \apjl, 723, L139

\bibitem[{{Brandt} {et~al.}(1988){Brandt}, {Scharmer}, {Ferguson}, {Shine}, \&
  {Tarbell}}]{Brandt:1988}
{Brandt}, P.~N., {Scharmer}, G.~B., {Ferguson}, S., {Shine}, R.~A., \&
  {Tarbell}, T.~D. 1988, \nat, 335, 238

\bibitem[{{David}(1961)}]{David:1961}
{David}, K.-H. 1961, \zap, 53, 37

\bibitem[{{de la Cruz Rodr{\'{\i}}guez} {et~al.}(2015){de la Cruz
  Rodr{\'{\i}}guez}, {L{\"o}fdahl}, {S{\"u}tterlin}, {Hillberg}, \& {Rouppe van
  der Voort}}]{Cruz:2015}
{de la Cruz Rodr{\'{\i}}guez}, J., {L{\"o}fdahl}, M.~G., {S{\"u}tterlin}, P.,
  {Hillberg}, T., \& {Rouppe van der Voort}, L. 2015, \aap, 573, A40

\bibitem[{{De Pontieu} {et~al.}(2014{\natexlab{a}}){De Pontieu}, {Rouppe van
  der Voort}, {McIntosh}, {Pereira}, {Carlsson}, {Hansteen}, {Skogsrud},
  {Lemen}, {Title}, {Boerner}, {Hurlburt}, {Tarbell}, {Wuelser}, {De Luca},
  {Golub}, {McKillop}, {Reeves}, {Saar}, {Testa}, {Tian}, {Kankelborg},
  {Jaeggli}, {Kleint}, \& {Martinez-Sykora}}]{DePontieu:2014b}
{De Pontieu}, B., {Rouppe van der Voort}, L., {McIntosh}, S.~W., {et~al.}
  2014{\natexlab{a}}, Science, 346, 1255732

\bibitem[{{De Pontieu} {et~al.}(2014{\natexlab{b}}){De Pontieu}, {Title},
  {Lemen}, {Kushner}, {Akin}, {Allard}, {Berger}, {Boerner}, {Cheung}, {Chou},
  {Drake}, {Duncan}, {Freeland}, {Heyman}, {Hoffman}, {Hurlburt}, {Lindgren},
  {Mathur}, {Rehse}, {Sabolish}, {Seguin}, {Schrijver}, {Tarbell},
  {W{\"u}lser}, {Wolfson}, {Yanari}, {Mudge}, {Nguyen-Phuc}, {Timmons}, {van
  Bezooijen}, {Weingrod}, {Brookner}, {Butcher}, {Dougherty}, {Eder},
  {Knagenhjelm}, {Larsen}, {Mansir}, {Phan}, {Boyle}, {Cheimets}, {DeLuca},
  {Golub}, {Gates}, {Hertz}, {McKillop}, {Park}, {Perry}, {Podgorski},
  {Reeves}, {Saar}, {Testa}, {Tian}, {Weber}, {Dunn}, {Eccles}, {Jaeggli},
  {Kankelborg}, {Mashburn}, {Pust}, {Springer}, {Carvalho}, {Kleint}, {Marmie},
  {Mazmanian}, {Pereira}, {Sawyer}, {Strong}, {Worden}, {Carlsson}, {Hansteen},
  {Leenaarts}, {Wiesmann}, {Aloise}, {Chu}, {Bush}, {Scherrer}, {Brekke},
  {Martinez-Sykora}, {Lites}, {McIntosh}, {Uitenbroek}, {Okamoto}, {Gummin},
  {Auker}, {Jerram}, {Pool}, \& {Waltham}}]{DePontieu:2014a}
{De Pontieu}, B., {Title}, A.~M., {Lemen}, J.~R., {et~al.} 2014{\natexlab{b}},
  \solphys, 289, 2733

\bibitem[{{Fedun} {et~al.}(2011){Fedun}, {Shelyag}, {Verth}, {Mathioudakis}, \&
  {Erd{\'e}lyi}}]{Fedun:2011}
{Fedun}, V., {Shelyag}, S., {Verth}, G., {Mathioudakis}, M., \& {Erd{\'e}lyi},
  R. 2011, Annales Geophysicae, 29, 1029

\bibitem[{{Kitiashvili} {et~al.}(2013){Kitiashvili}, {Kosovichev}, {Lele},
  {Mansour}, \& {Wray}}]{Kitiashvili:2013}
{Kitiashvili}, I.~N., {Kosovichev}, A.~G., {Lele}, S.~K., {Mansour}, N.~N., \&
  {Wray}, A.~A. 2013, \apj, 770, 37

\bibitem[{{Kitiashvili} {et~al.}(2012{\natexlab{a}}){Kitiashvili},
  {Kosovichev}, {Mansour}, {Lele}, \& {Wray}}]{Kitiashvili:2012a}
{Kitiashvili}, I.~N., {Kosovichev}, A.~G., {Mansour}, N.~N., {Lele}, S.~K., \&
  {Wray}, A.~A. 2012{\natexlab{a}}, \physscr, 86, 018403

\bibitem[{{Kitiashvili} {et~al.}(2011){Kitiashvili}, {Kosovichev}, {Mansour},
  \& {Wray}}]{Kitiashvili:2011}
{Kitiashvili}, I.~N., {Kosovichev}, A.~G., {Mansour}, N.~N., \& {Wray}, A.~A.
  2011, \apjl, 727, L50

\bibitem[{{Kitiashvili} {et~al.}(2012{\natexlab{b}}){Kitiashvili},
  {Kosovichev}, {Mansour}, \& {Wray}}]{Kitiashvili:2012b}
{Kitiashvili}, I.~N., {Kosovichev}, A.~G., {Mansour}, N.~N., \& {Wray}, A.~A.
  2012{\natexlab{b}}, \apjl, 751, L21

\bibitem[{{Leenaarts} {et~al.}(2013){Leenaarts}, {Pereira}, {Carlsson},
  {Uitenbroek}, \& {De Pontieu}}]{Leenaarts:2013}
{Leenaarts}, J., {Pereira}, T.~M.~D., {Carlsson}, M., {Uitenbroek}, H., \& {De
  Pontieu}, B. 2013, \apj, 772, 90

\bibitem[{{Lemen} {et~al.}(2012){Lemen}, {Title}, {Akin}, {Boerner}, {Chou},
  {Drake}, {Duncan}, {Edwards}, {Friedlaender}, {Heyman}, {Hurlburt}, {Katz},
  {Kushner}, {Levay}, {Lindgren}, {Mathur}, {McFeaters}, {Mitchell}, {Rehse},
  {Schrijver}, {Springer}, {Stern}, {Tarbell}, {Wuelser}, {Wolfson}, {Yanari},
  {Bookbinder}, {Cheimets}, {Caldwell}, {Deluca}, {Gates}, {Golub}, {Park},
  {Podgorski}, {Bush}, {Scherrer}, {Gummin}, {Smith}, {Auker}, {Jerram},
  {Pool}, {Soufli}, {Windt}, {Beardsley}, {Clapp}, {Lang}, \&
  {Waltham}}]{Lemen:2012}
{Lemen}, J.~R., {Title}, A.~M., {Akin}, D.~J., {et~al.} 2012, \solphys, 275, 17

\bibitem[{{Manso Sainz} {et~al.}(2011){Manso Sainz}, {Mart{\'{\i}}nez
  Gonz{\'a}lez}, \& {Asensio Ramos}}]{Manso:2011}
{Manso Sainz}, R., {Mart{\'{\i}}nez Gonz{\'a}lez}, M.~J., \& {Asensio Ramos},
  A. 2011, \aap, 531, L9

\bibitem[{{Moll} {et~al.}(2011){Moll}, {Cameron}, \&
  {Sch{\"u}ssler}}]{Moll:2011}
{Moll}, R., {Cameron}, R.~H., \& {Sch{\"u}ssler}, M. 2011, \aap, 533, A126

\bibitem[{{Pereira} {et~al.}(2015){Pereira}, {Carlsson}, {De Pontieu}, \&
  {Hansteen}}]{Pereira:2015}
{Pereira}, T.~M.~D., {Carlsson}, M., {De Pontieu}, B., \& {Hansteen}, V. 2015,
  \apj, 806, 14

\bibitem[{{Pesnell} {et~al.}(2012){Pesnell}, {Thompson}, \&
  {Chamberlin}}]{Pesnell:2012}
{Pesnell}, W.~D., {Thompson}, B.~J., \& {Chamberlin}, P.~C. 2012, \solphys,
  275, 3

\bibitem[{{Rouppe van der Voort} {et~al.}(2015){Rouppe van der Voort}, {De
  Pontieu}, {Pereira}, {Carlsson}, \& {Hansteen}}]{Voort:2015}
{Rouppe van der Voort}, L., {De Pontieu}, B., {Pereira}, T.~M.~D., {Carlsson},
  M., \& {Hansteen}, V. 2015, \apjl, 799, L3

\bibitem[{{Scharmer} {et~al.}(2003{\natexlab{a}}){Scharmer}, {Bjelksjo},
  {Korhonen}, {Lindberg}, \& {Petterson}}]{Scharmer:2003a}
{Scharmer}, G.~B., {Bjelksjo}, K., {Korhonen}, T.~K., {Lindberg}, B., \&
  {Petterson}, B. 2003{\natexlab{a}}, in Society of Photo-Optical
  Instrumentation Engineers (SPIE) Conference Series, Vol. 4853, Innovative
  Telescopes and Instrumentation for Solar Astrophysics, ed. S.~L. {Keil} \&
  S.~V. {Avakyan}, 341--350

\bibitem[{{Scharmer} {et~al.}(2003{\natexlab{b}}){Scharmer}, {Dettori},
  {Lofdahl}, \& {Shand}}]{Scharmer:2003b}
{Scharmer}, G.~B., {Dettori}, P.~M., {Lofdahl}, M.~G., \& {Shand}, M.
  2003{\natexlab{b}}, in Society of Photo-Optical Instrumentation Engineers
  (SPIE) Conference Series, Vol. 4853, Innovative Telescopes and
  Instrumentation for Solar Astrophysics, ed. S.~L. {Keil} \& S.~V. {Avakyan},
  370--380

\bibitem[{{Scharmer} {et~al.}(2008){Scharmer}, {Narayan}, {Hillberg}, {de la
  Cruz Rodr{\'{\i}}guez}, {L{\"o}fdahl}, {Kiselman}, {S{\"u}tterlin}, {van
  Noort}, \& {Lagg}}]{Scharmer:2008}
{Scharmer}, G.~B., {Narayan}, G., {Hillberg}, T., {et~al.} 2008, \apjl, 689,
  L69

\bibitem[{{Scherrer} {et~al.}(2012){Scherrer}, {Schou}, {Bush}, {Kosovichev},
  {Bogart}, {Hoeksema}, {Liu}, {Duvall}, {Zhao}, {Title}, {Schrijver},
  {Tarbell}, \& {Tomczyk}}]{Scherrer:2012}
{Scherrer}, P.~H., {Schou}, J., {Bush}, R.~I., {et~al.} 2012, \solphys, 275,
  207

\bibitem[{{Shelyag} {et~al.}(2011){Shelyag}, {Keys}, {Mathioudakis}, \&
  {Keenan}}]{Shelyag:2011}
{Shelyag}, S., {Keys}, P., {Mathioudakis}, M., \& {Keenan}, F.~P. 2011, \aap,
  526, A5

\bibitem[{{Stein} \& {Nordlund}(2000)}]{Stein:2000}
{Stein}, R.~F. \& {Nordlund}, {\AA}. 2000, \solphys, 192, 91

\bibitem[{{Sturrock} \& {Uchida}(1981)}]{Sturrock:1981}
{Sturrock}, P.~A. \& {Uchida}, Y. 1981, \apj, 246, 331

\bibitem[{{Tsiropoula}(2000)}]{Tsiropoula:2000}
{Tsiropoula}, G. 2000, \na, 5, 1

\bibitem[{{van Noort} {et~al.}(2005){van Noort}, {Rouppe van der Voort}, \&
  {L{\"o}fdahl}}]{Noort:2005}
{van Noort}, M., {Rouppe van der Voort}, L., \& {L{\"o}fdahl}, M.~G. 2005,
  \solphys, 228, 191

\bibitem[{{Vargas Dom{\'{\i}}nguez} {et~al.}(2011){Vargas Dom{\'{\i}}nguez},
  {Palacios}, {Balmaceda}, {Cabello}, \& {Domingo}}]{Vargas:2011}
{Vargas Dom{\'{\i}}nguez}, S., {Palacios}, J., {Balmaceda}, L., {Cabello}, I.,
  \& {Domingo}, V. 2011, \mnras, 416, 148

\bibitem[{{Wedemeyer} {et~al.}(2013){Wedemeyer}, {Scullion}, {Steiner}, {de la
  Cruz Rodriguez}, \& {Rouppe van der Voort}}]{Wedemeyer:2013}
{Wedemeyer}, S., {Scullion}, E., {Steiner}, O., {de la Cruz Rodriguez}, J., \&
  {Rouppe van der Voort}, L.~H.~M. 2013, Journal of Physics Conference Series,
  440, 012005

\bibitem[{{Wedemeyer} \& {Steiner}(2014)}]{Wedemeyer:2014}
{Wedemeyer}, S. \& {Steiner}, O. 2014, \pasj, 66, 10

\bibitem[{{Wedemeyer-B{\"o}hm} \& {Rouppe van der
  Voort}(2009)}]{Wedemeyer:2009}
{Wedemeyer-B{\"o}hm}, S. \& {Rouppe van der Voort}, L. 2009, \aap, 507, L9

\bibitem[{{Wedemeyer-B{\"o}hm} {et~al.}(2012){Wedemeyer-B{\"o}hm}, {Scullion},
  {Steiner}, {Rouppe van der Voort}, {de La Cruz Rodriguez}, {Fedun}, \&
  {Erd{\'e}lyi}}]{Wedemeyer:2012}
{Wedemeyer-B{\"o}hm}, S., {Scullion}, E., {Steiner}, O., {et~al.} 2012, \nat,
  486, 505

\bibitem[{{Zirker}(1993)}]{Zirker:1993}
{Zirker}, J.~B. 1993, \solphys, 147, 47

\end{thebibliography}

\end{document}